\documentclass[aps,epsfig,twocolumn,showpacs,email,amssymb,nobibnotes]{revtex4}
\usepackage{graphicx,amsmath,amssymb,psfrag,epsfig,multirow,slashbox}
\begin{document}
\title{Stochastic oscillations in models of epidemics on a network of cities}

\author{G.~Rozhnova$^1$, A.~Nunes$^1$ and A.~J.~McKane$^{1,2}$}

\affiliation{$^1$Centro de F{\'\i}sica da Mat{\'e}ria Condensada and 
Departamento de F{\'\i}sica, Faculdade de Ci{\^e}ncias da Universidade de 
Lisboa, P-1649-003 Lisboa Codex, Portugal \\
$^2$Theoretical Physics Division, School of Physics and Astronomy,
University of Manchester, Manchester M13 9PL, United Kingdom
}

\begin{abstract}
We carry out an analytic investigation of stochastic oscillations in a
susceptible-infected-recovered model of disease spread on a network of $n$
cities. In the model a fraction $f_{jk}$ of individuals from city $k$ commute
to city $j$, where they may infect, or be infected by, others. Starting from
a continuous time Markov description of the model the deterministic equations,
which are valid in the limit when the population of each city is infinite, are
recovered. The stochastic fluctuations about the fixed point of these
equations are derived by use of the van Kampen system-size expansion. The fixed
point structure of the deterministic equations is remarkably simple: a unique
non-trivial fixed point always exists and has the feature that the fraction of
susceptible, infected and recovered individuals is the same for each city
irrespective of its size. We find that the stochastic fluctuations have an
analogously simple dynamics: all oscillations have a single frequency, equal
to that found in the one city case. We interpret this phenomenon in terms of 
the properties of the spectrum of the matrix of the linear approximation of  
the deterministic equations at the fixed point.
\end{abstract}

\maketitle

\section{Introduction}
\label{intro}
Two of the ideas that are currently dominating the discussion of modeling 
epidemic spread are those of stochasticity and network structure 
\cite{pastor1,MattJKeeling09222005,PhysRevLett.96.208701,volz,colizza,STdG07}. 
Deterministic models of the Susceptible-Infected-Recovered (SIR) type have a 
long history \cite{and_may,KR07} and have been thoroughly investigated 
\cite{Sch85,KeelingGrenfellMeasles} along with many extensions of the models 
such as age classes or higher-order nonlinear interaction terms. Although 
stochasticity, due to random processes at the level of individuals, and 
networks, either between individuals or towns and cities, were recognised 
early on as significant and important factors, the tendency was to model them 
through computer simulations. This is not surprising: it is rather 
straightforward to deal with stochastic behavior in simulations, and similarly 
the analytic methods available to investigate complex networks, especially 
adaptive networks, are limited. There has also been a tendency towards 
developing extremely detailed agent-based models to study disease spread  
\cite{minorityhealth1293,Meyers05networktheory,grenfellphysA2010,Aleman:2011:NAS:2000906.2000914}, which are the antithesis of the simple analytic approach 
based on the original SIR deterministic model.

In parallel with these developments, however, there have been several efforts
to extend analytic studies into the realm of stochastic and network dynamics.
The SIR model can be formulated as an individual-based model (IBM) which can
form a starting point for both an analytical treatment, based on the master
equation (continuous-time Markov chain) \cite{van_kampen,Risken}, and numerical
simulations, based on the Gillespie algorithm \cite{gillespie}. The analytical 
studies use the system-size expansion of van Kampen to reduce the master 
equation to the set of deterministic equations previously studied, together 
with a set of stochastic differential equations for the deviations from the 
deterministic result. As long as one is not too close to the fade-out boundary,
there is no need to go beyond next-to-leading order in the expansion parameter, 
$1/\sqrt{N}$, where $N$ is the number of individuals in the system. This
already gives results which are, in general, in almost perfect agreement with
the results of simulations \cite{mckane_new2,alonso}.

This approach has been used to study the stochastic version of the standard 
SIR model \cite{alonso}, the Susceptible-Exposed-Infected-Recovered (SEIR) 
model \cite{seasonalRN}, both these models with annual forcing 
\cite{seasonalRN,epidforcingmckane}, staged-models 
\cite{parisi,stagedmodelAndrew}, the pair-approximation in networked models 
\cite{PhysRevE.79.041922,PhysRevE.80.051915}, amongst others. In this
paper we extend the treatment to a metapopulation model for disease spread, 
which consists of $n$ cities (labeled $j=1,\ldots,n$), each of which contains 
$N_j$ individuals. A fraction of the population of city $k$, $f_{jk}$, commutes
to city $j$ and this defines the strength of the link from node $k$ to $j$ 
in the network of cities. We will show that the methods used in the case of 
one city carry over to the case where the system comprises of a network of 
cities, and that a surprisingly simple set of results can be derived which 
allow us to make quite general predictions for a class of stochastic networked 
models of epidemics.

The starting point for our analysis is a specification of how commuters move
between cities in the network. As will become clear, the model we arrive 
at will not depend on the details of how and when these exchanges take place.
We then write down transition rates for the usual SIR process, now taking 
account the city of origin of the infector and infected individuals. From
the resulting equation we can immediately find the deterministic equations 
corresponding to the stochastic model when $N_j \to \infty$ for all $j$. 
Deterministic models of this type began to be considered long ago 
\cite{gonorrhea_yorke} and the existence and stability properties of the 
endemic equilibrium were studied for different formulations of the coupling 
between the cities and of the disease dynamics 
\cite{lloyd_may,mech_approach,multi_city}. Stochastic effects in these systems 
have also been analyzed from the point of view of the relation between spatial 
heterogeneity, disease extinction and the threshold for disease onset 
\cite{lloyd_may,Hagenaars-etal2004,Colizza_Vespignani_2008,bathelemy2010_etal}. 

Some rather strong and general results on the uniqueness and global stability 
of the fixed points of the deterministic model are known \cite{guo-etal2008}.
We will use these results and then go beyond this leading-order analysis to 
determine the linear stochastic corrections that characterize the 
quasi-stationary state of the finite system. As expected, the qualitative 
predictions of the deterministic model are shown to be incorrect; instead 
large stochastic cycles are found, although their form is much simpler than 
might naively have been expected. We show that this is, in part, a reflection 
of the special nature of the fixed points of the deterministic model.

The outline of the paper is as follows. In Section~\ref{two_city} we describe 
the basic model and apply it to the case of two cities. The generalization 
to the $n$-city case in given in Section~\ref{n_city}. The results for the 
form of the sustained oscillations are given in Section~\ref{results} and 
we conclude in Section~\ref{conclude}. Two appendices contain technical 
details which are too cumbersome to include in the main text.

\section{Two-city model}
\label{two_city}
In this section we will formulate the model when there are only two cities; the
general $n$ city case described in Section~\ref{n_city} does not introduce 
any new points of principle and is easily explained once the two city case 
has been understood.

The SIR model consists of three classes of individuals: susceptibles, infected
and recovered. The number of individuals in the three classes belonging to 
city $j$ will be denoted by $S_{j}, I_{j}$ and $R_j$ respectively. We assume 
that births and deaths are coupled at the individual level, so that when an 
individual dies another (susceptible) individual is born. This means that
the number of individuals belonging to any one city, $N_j$, does not change 
with time, and so the number of recovered individuals is not an independent 
variable: $R_{j}=N_{j}-S_{j}-I_{j}$, where $j=1,2$ \cite{alonso}.

There are four processes in the SIR model which cause transitions to a new 
state: infection, recovery, death of an infected individual and death of a 
recovered individual. The death of a susceptible individual does not cause
a transition, since it is immediately replaced by another, newborn, individual
which is by definition susceptible. Of the four listed processes, the final
three only involve one individual and so only involve one city. The transition
rates are \cite{alonso}:
\begin{itemize}
\item[(a)] Recovery of an infective individual (and creation of a recovered
individual)
\begin{align}
T_{1} & \equiv T(S_1,I_{1}-1,S_2,I_2|S_1,I_1,S_2,I_2) = \gamma I_1 , \nonumber \\
T_{2} & \equiv T(S_1,I_1,S_2,I_{2}-1|S_1,I_1,S_2,I_2) = \gamma I_2 .
\label{recovery_2}
\end{align}
\item[(b)] Death of an infected individual (and birth of a susceptible 
individual): 
\begin{align}
T_{3} & \equiv T(S_{1}+1,I_{1}-1,S_2,I_2|S_1,I_1,S_2,I_2) = \mu I_1 , \nonumber \\
T_{4} & \equiv T(S_1,I_1,S_{2}+1,I_{2}-1|S_1,I_1,S_2,I_2) = \mu I_2 .
\label{death_infect_2}
\end{align}
\item[(c)] Death of a recovered individual (and birth of a susceptible 
individual): 
\begin{align}
T_{5} & \equiv T(S_{1}+1,I_1,S_2,I_2|S_1,I_1,S_2,I_2) = \mu (N_{1}-S_{1}-I_{1}), 
\nonumber \\
T_{6} & \equiv T(S_1,I_1,S_{2}+1,I_2|S_1,I_1,S_2,I_2) = \mu (N_{2}-S_{2}-I_{2}).
\nonumber \\
\label{death_recover_2}
\end{align}
\end{itemize}
Here $\gamma$ and $\mu$ are parameters which respectively characterize the 
rate of recovery and of birth/death. 

The infection processes introduce the role of the commuters. We let $f_{21}$ 
be the fraction of the population from city $1$ which commutes to city $2$,
leaving a fraction $(1-f_{21})$ of the population as residents of city $1$.
Similarly, for commuters from city $2$, as illustrated in Fig.~$1$. We note
that the number of individuals in city $j$ is 
$M_{j}=(1-f_{kj})N_{j}+f_{jk}N_{k}$, where $j \neq k$. We will not specify the
nature of the commute in more detail and assume that the $f_{jk}$ are a 
property of the corresponding pair of cities that defines the overall average 
fraction of time that an individual from one city spends in the other city. 
These coefficients will be taken as a coarse-grained measure of the 
demographic coupling  between the cities that will be applied to all 
individuals independently of disease status and do not discriminate between 
different types of stays with their typical frequencies and durations.

To see the nature of the infective interactions that occur, we first fix our
attention on those involving susceptible individuals from city $1$. There
are four types of term:
\begin{itemize}
\item[(i)] Infective residents in city $1$ infect susceptible residents in
city $1$.
 \item[(ii)] Infective commuters from city $2$ infect susceptible residents in
city $1$.
\item[(iii)] Infective residents in city $2$ infect susceptible commuters from
city $1$.
\item[(iv)] Infective commuters from city $1$ infect susceptible commuters 
from city $1$ in city $2$.
\end{itemize}
The rates for these to occur according to the usual prescription for the SIR
model \cite{alonso} are:
\begin{itemize}
\item[(i)] $\beta\,\left( 1 - f_{21}\right)S_1\,\left( 1 - f_{21}\right)I_1/M_1$,
\item[(ii)] $\beta\,\left( 1 - f_{21}\right)S_1\,f_{12}I_2/M_1$,
\item[(iii)] $\beta\,f_{21}S_1\,\left( 1 - f_{12}\right)I_2/M_2$,
\item[(iv)] $\beta\,f_{21}S_1\,f_{21}I_1/M_2$,
\end{itemize}
where $\beta$ is the parameter which sets the overall rate of infection. Adding
these rates together we obtain the total transition rate for infection of 
$S_1$ individuals as 
\begin{displaymath}
\beta \left[ c_{11} \frac{S_1 I_1}{N_1} + c_{12} \frac{S_1 I_2}{N_2} \right],
\end{displaymath}
where 
\begin{eqnarray}
c_{11} &=& \frac{\left( 1 - f_{21}\right)^{2} N_1}{M_1} +
\frac{f^{2}_{21} N_{1}}{M_2}, \nonumber \\
c_{12} &=& \frac{\left( 1 - f_{21}\right)f_{12}N_2}{M_1} +
\frac{f_{21} \left( 1 - f_{12}\right)N_2}{M_2}. \nonumber
\end{eqnarray}

A similar analysis can be made for the transitions involving susceptible 
individuals from city 2. Putting these results together we obtain the 
transition rates for infection as
\begin{itemize}
\item[(d)] Infection of a susceptible individual:
\begin{align}
T_{7} &\equiv T(S_{1}-1,I_{1}+1,S_2,I_2|S_1,I_1,S_2,I_2) \nonumber \\
&= \beta \left[ c_{11} \frac{S_1 I_1}{N_1} + c_{12} \frac{S_1 I_2}{N_2} \right],
\nonumber \\
T_{8} &\equiv T(S_1,I_1,S_{2}-1,I_{2}+1|S_1,I_1,S_2,I_2) \nonumber \\
&=  \beta \left[ c_{21} \frac{S_2 I_1}{N_1} + c_{22} \frac{S_2 I_2}{N_2} \right],
\label{infection_2}
\end{align}
\end{itemize}
where
\begin{eqnarray}
c_{11}&=&\dfrac{(1-f_{21})^2}{1-f_{21}+f_{12}q}+
\dfrac{f_{21}^2}{f_{21}+(1-f_{12})q},\nonumber \\
c_{12}&=&\dfrac{(1-f_{21})f_{12}q}{1-f_{21}+f_{12}q}+
\dfrac{f_{21}(1-f_{12})q}{f_{21}+(1-f_{12})q},\nonumber \\
c_{21}&=&\dfrac{(1-f_{12})f_{21}}{f_{21}+(1-f_{12})q}+
\dfrac{f_{12}(1-f_{21})}{1-f_{21}+f_{12}q},\nonumber \\
c_{22}&=&\dfrac{(1-f_{12})^2q}{f_{21}+(1-f_{12})q}+
\dfrac{f_{12}^2q}{1-f_{21}+f_{12}q},
\label{cs_2}
\end{eqnarray}
and $q=N_2/N_1$. We assume that $N_1$ and $N_2$ are not too different, so
that $q$ is neither very small nor very large.


\begin{figure}
\centering
\includegraphics[width=0.6\columnwidth]{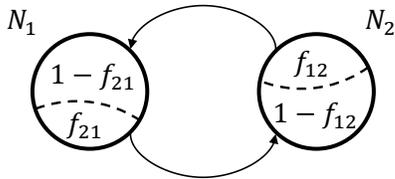}
{\caption{A fraction $f_{jk}$ of residents of city $k$ commute to city $j$,
where $j,k=1,2$.}}
\label{fig1}
\end{figure}


The model is defined by the transitions rates in 
Eqs.~(\ref{recovery_2})-(\ref{infection_2}). It is interesting that the 
transitions due to infection depend on the fractions $f_{jk}$ only through
the constants $c_{jk}$ defined in Eq.~(\ref{cs_2}). Other ways of accounting 
for commuting individuals would typically still give rise to the form given in
Eq.~(\ref{infection_2}), but with the constants $c_{jk}$ defined in a 
different way.  

Since our counting of the ways that infection takes place was exhaustive, we
expect that the constants $c_{jk}$ are not independent. It is straightforward 
to check that they obey the following relations:
\begin{equation}
c_{11}+c_{12}=1,\ \ c_{21}+c_{22}=1, \ \ c_{12}=c_{21}q.
\label{relations_2}
\end{equation}
So there are only two independent parameters in additional to the usual SIR
parameters $\beta, \gamma$ and $\mu$ found in the single city case, and we
choose these to be $c_{12}$ and $q=N_2/N_1$.
Our results will be given in terms of these two parameters.
It is easy to see that, for each $q$, the range of $c_{12}$ is the interval
$[0, q/(q+1)] $ where the maximum is attained for $f_{21} + f_{12} = 1$.
While exploring the general behavior of the system we will consider the 
$c_{jk}$ independently of the underlying microscopic model as positive 
parameters that take values in the  wider admissible range defined by the 
constraints (\ref{relations_2}). 
 
Having specified the model it may be investigated in two ways as indicated 
in Section~\ref{intro}. First, it can simulated with Gillespie's algorithm 
\cite{gillespie}, or some equivalent method. Second, it can be studied 
analytically by constructing the master equation and performing van Kampen's 
system size expansion on this equation. This will be the main focus of this 
paper. For notational convenience we will denote the states of the system by 
$\sigma \equiv \{S_1,I_1,S_2,I_2\}$, recalling that the number of recovered 
individuals from each city may be written in terms of these variables. The 
master equation gives the time evolution of $P(\sigma,t)$, the probability 
distribution for finding the system in state $\sigma$ at time $t$. It takes 
the form \cite{van_kampen,Risken}
\begin{equation}
\frac{dP(\sigma,t)}{dt} = \sum_{\sigma' \neq \sigma} \sum^{8}_{a=1}
\left[ T_{a}(\sigma | \sigma')P(\sigma',t) - 
T_{a}(\sigma' | \sigma)P(\sigma,t) \right],
\label{master}
\end{equation}
where $T_{a}(\sigma | \sigma')$, $a=1,\ldots,8$ are the transition rates from 
the state $\sigma'$ to the state $\sigma$ given explicitly in 
Eqs.~(\ref{recovery_2})-(\ref{infection_2}).

The full master equation (\ref{master}) cannot be solved, but the van Kampen 
system-size expansion when taken to next-to-leading order usually gives results
which are in excellent agreement with simulations. We will see that this will
also be the case in the extensions of the method which we are exploring in this
paper. The system-size expansion starts by making the following ansatz 
\cite{van_kampen}:
\begin{equation}
S_j = N_js_j+N_j^{1/2}x_j,\ \ I_j=N_ji_j+N_j^{1/2}y_j,
\label{ansatz}
\end{equation}
where $j=1,2$. Here $s_j = \lim_{N_j \to \infty} S_j/N_j$ and 
$i_j = \lim_{N_j \to \infty} I_j/N_j$ are the fraction of individuals from city
$j$ which are respectively susceptible and infected in the deterministic 
limit. The quantities $x_j$ and $y_j$ are the stochastic deviations from these
deterministic results, suitably scaled so that they also become continuous in
the limit of large population sizes. The ansatz (\ref{ansatz}) is substituted 
into Eq.~(\ref{master}) and powers of $\sqrt{N_j}$ on the left- and right-hand 
sides matched up. The leading order contribution gives the deterministic 
equations of the model and the next-to-leading order linear stochastic 
differential equations for $x_j$ and $y_j$. We shall not describe the method 
in great detail, since it is described clearly in van Kampen's book 
\cite{van_kampen} and in many papers, including several on the SIR model 
\cite{alonso,STdG07,parisi}. Instead we will outline the main results of the 
approximation in the remainder of this section, and give some explicit 
intermediate formulas in Appendix~\ref{app:a}.

The deterministic equations which are found to first order in the system-size
expansion can also be obtained by multiplying Eq.~(\ref{master}) by 
$S_1,I_1,S_2$ and $I_2$ in turn and then summing over all states $\sigma$. 
This yields
\begin{eqnarray}
\dfrac{ds_1}{dt}&=& -\beta s_1 \left(c_{11}i_1+c_{12}i_2\right) + \mu(1-s_1),
\nonumber \\
\dfrac{ds_2}{dt}&=& -\beta s_2 \left(c_{21}i_1+c_{22}i_2\right) + \mu(1-s_2),
\nonumber \\
\dfrac{di_1}{dt}&=&\beta s_1 \left(c_{11}i_1+c_{12}i_2\right)-(\gamma+\mu)i_1,
\nonumber \\
\dfrac{di_2}{dt}&=&\beta s_2 \left(c_{21}i_1+c_{22}i_2\right)-(\gamma+\mu)i_2.
\label{deter_2}
\end{eqnarray}
For the case of cities with equal population sizes, these have been previously 
found and analyzed in \cite{mech_approach}. In the 
context of the present work we are mainly interested in the fixed points of
these equations. We will not discuss these here, instead we will wait until
Section~\ref{n_city}, where the case of $n$ cities will be discussed when we 
can give a more general treatment.

Of more interest to us in this paper are the variables $x_j$ and $y_j$ which
describe the linear fluctuations around trajectories of the deterministic 
set of equations (\ref{deter_2}). For convenience we will introduce the vector
of these fluctuations $\mathbf{z}=(x_1,x_2,y_1,y_2)$. Our focus will be 
on fluctuations in the stationary state, that is, about the non-trivial 
fixed point of the deterministic equations (which we will show in the next
section is unique). The fluctuations obtained through the system-size 
expansion obey a linear Fokker-Planck equation, which is equivalent to a set 
of stochastic differential equations of the form \cite{Risken}
\begin{equation}
\frac{d z_J}{dt} = \sum_{K=1}^4 A_{JK} z_K + \eta_J(t), \ \ J=1,\ldots,4,
\label{Langevin}
\end{equation}
where $\eta_J(t)$ are Gaussian noise terms with zero mean which satisfy 
$\langle\eta_J(t)\eta_K(t')\rangle = B_{JK}\delta(t-t')$. Since the fluctuations
are about the fixed point, the $4 \times 4$ matrices $A$ and $B$ are 
independent of time, and completely characterize the fluctuations. They are
given explicitly in Appendix~\ref{app:a}.

The fluctuations will be analyzed in detail in Section~\ref{results}, when 
they will also be compared to the results of numerical simulations. Before 
discussing this, we will generalize the discussion of this section to an 
arbitrary network of $n$ cities.

\section{Arbitrary network structure}
\label{n_city}
In this section we will generalize the content of Section~\ref{two_city} to
$n$ cities and also find the fixed points of the deterministic dynamics in 
this case.


\begin{figure}
\centering
\includegraphics[width=0.6\columnwidth]{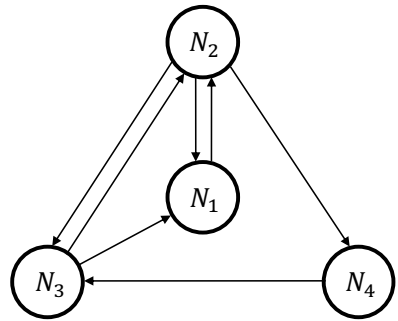}
{\caption{Individuals commute between $n$ cities, illustrated for a particular
network when $n=4$.}}
\label{fig2}
\end{figure}


\subsection{$n$-city model}
\label{subsec:n_city}
We use the same notation as in Section~\ref{two_city}, labeling the cities 
by $j$ and $k$ which now run from $1$ to $n$. It will be convenient to 
introduce the quantity
\begin{equation}
f_j = \sum_{k\neq j} f_{kj},
\label{f_j}
\end{equation}
so that the number of individuals in city $j$ may be written as
\begin{eqnarray}
M_j &=& \Bigl[ 1 - \sum_{k\neq j} f_{kj} \Bigr]N_j + \sum_{k\neq j} f_{jk}N_k 
\nonumber \\
&=& \left( 1 - f_j \right)N_j + \sum_{k\neq j} f_{jk}N_{k}.
\label{M_n}
\end{eqnarray}

There are, once again, four types of term in the process of infection (see 
Figure~2) and we again fix our attention on those involving susceptible 
individuals from city 1:
\begin{itemize}
\item[(i)] Infective residents in city $1$ infect susceptible residents in 
city $1$. This gives a rate of $\beta (1-f_1) S_1 (1-f_1)I_1/M_1$.
\item[(ii)] Infective commuters from city $j$, $j=2,\ldots,n$, infect 
susceptible residents in city $1$. This gives a rate, summing over all $j$, of
$\beta (1-f_1)S_1 \sum_{j\neq 1} f_{1j}I_j/M_1$.
\item[(iii)] Infective residents in city $j$, $j=2,\ldots,n$, infect 
susceptible commuters from city $1$. This gives a rate, summing over all $j$, of
$\beta \sum_{j\neq 1} (1-f_j)I_j f_{j1}S_1/M_j$.
\item[(iv)] Infective commuters from city $k$ (including city $1$) infect 
susceptible commuters from city $1$ in city $j$. This gives a total rate of
$\beta \sum_{j\neq1} f_{j1}S_1 \sum_{k\neq j}f_{jk} I_k/M_j$.
\end{itemize}
Since the transition rates for recovery and birth/death are simple extensions
of those for two cities we can now write down the transition rates for the
$n$-city model as:
\begin{itemize}
\item[(a)] Recovery of an infective individual (and creation of a recovered
individual)
\begin{equation}
T_{j} \equiv T(S_1,I_1,\ldots,S_j,I_{j}-1,\ldots,S_n,I_n|\sigma) = \gamma I_{j},
\label{recovery_n}
\end{equation}
\item[(b)] Death of an infected individual (and birth of a susceptible 
individual): 
\begin{equation}
T_{n+j} \equiv 
T(S_1,I_1,\ldots,S_{j}+1,I_{j}-1,\ldots,S_n,I_n|\sigma) = \mu I_{j},
\label{death_infect_n}
\end{equation}
\item[(c)] Death of a recovered individual (and birth of a susceptible 
individual): 
\begin{align}
T_{2n+j} &\equiv T(S_1,I_1,\ldots,S_{j}+1,I_j\ldots,S_n,I_n|\sigma) \nonumber \\
&= \mu (N_j - S_j -I_j),
\label{death_recover_n}
\end{align}
\item[(d)] Infection of a susceptible individual:
\begin{align}
T_{3n+j} &\equiv T(S_1,I_1,\ldots,S_{j}-1,I_{j}+1\ldots,S_n,I_n|\sigma) 
\nonumber \\
&= \beta \sum^{n}_{k=1} c_{jk} \frac{S_j I_k}{N_k},
\label{infection_n}
\end{align}
\end{itemize}
where $\sigma\equiv\{S_1,I_1,\ldots,S_j,I_j\ldots,S_n,I_n\}$ and where 
$j=1,\ldots,n$. The coefficients $c_{jk}$ in Eq.~(\ref{infection_n}) may be
read off from the terms (i)-(iv), but they are sufficiently complicated to 
write down in full that we only list them in Appendix~\ref{app:b}. In that 
Appendix we also show that relations between the $c_{jk}$, analogous to those
given in Eq.~({\ref{relations_2}) for the two-city case hold, and are given by
\begin{equation}
c_{jj} + \sum_{k\neq j}c_{jk} = 1; \ \ 
c_{kj} = \left( \frac{N_j}{N_k}\right) c_{jk}; \ j,k=1,\ldots,n.
\label{relations_n}
\end{equation}
So in the $n$-city model, there are $n(n-1)/2$ independent coupling parameters 
$c_{jk}$ and $(n-1)$ parameters for city sizes in additional to the usual 
epidemiological parameters. Note that if all city sizes are equal the second 
relation in Eq.~(\ref{relations_n}) reduces to $c_{kj}=c_{jk}$. This symmetry 
will be used in the subsequent analysis. 

Following the same path as in Section~\ref{two_city}, having specified the 
model by giving the transition rates, we move on to the dynamics. The process
is Markovian and so satisfies the master equation (\ref{master}) except now
the sum on $a$ goes from $1$ to $4n$. As detailed in Appendix~\ref{app:a},
invoking the van Kampen ansatz (\ref{ansatz}) gives the following deterministic
equations to leading order:
\begin{eqnarray}
\dfrac{ds_j}{dt}&=& -\beta s_j \sum^{n}_{k=1} c_{jk} i_k + 
\mu \left( 1 - s_j \right), \nonumber \\
\dfrac{di_j}{dt}&=& \beta s_j \sum^{n}_{k=1} c_{jk} i_k -
\left( \gamma + \mu \right) i_j,
\label{deter_n}
\end{eqnarray}
where $j=1,\ldots,n$. At next-to-leading order the fluctuations are found to 
satisfy the linear stochastic differential equation (\ref{Langevin}), but with 
$J,K=1,\ldots,2n$. The two matrices $A$ and $B$ are given explicitly in
Appendix~\ref{app:a}. They are independent of time, since they are evaluated 
at the fixed points of the dynamics (\ref{deter_n}). For the rest of this 
section we will investigate the fixed point structure of these equations.

\subsection{The fixed points of the deterministic equations}
\label{fixed_points}
The fixed points of the deterministic equations (\ref{deter_n}) will be 
denoted by asterisks. Adding the two sets of equations we immediately see 
that
\begin{equation}
\left( \gamma + \mu \right)i^{*}_j = \mu\left( 1 - s^{*}_j \right), \ \ 
j=1,\ldots,n.
\label{first_FP_eqn}
\end{equation}
Using this equation to eliminate the $i^{*}_{j}$, and also using 
Eq.~(\ref{relations_n}), one finds that
\begin{equation}
s^{*}_j \left[ \left( \beta + \gamma + \mu \right) -
\beta\sum^n_{k=1}c_{jk}s^{*}_k \right] = \left( \gamma + \mu \right), \ \ 
j=1,\ldots,n.
\label{second_FP_eqn}
\end{equation}

Two fixed points can be found by inspection. First, suppose one of the 
$i^{*}_{j}$ is zero, for instance $i^{*}_{\ell} = 0$. Then from 
Eq.~(\ref{first_FP_eqn}) $s^{*}_{\ell} = 1$. From Eq.~(\ref{deter_n}) we see 
immediately that $\sum^{n}_{k=1} c_{\ell k} i^{*}_k = 0$. Since the coefficients 
$c_{jk}$ are non-negative (see Appendix~\ref{app:b}), then $i^{*}_k = 0$ for 
all $k$ as long as $c_{\ell k} \neq 0$. Using the $i^{*}_k$ which are zero as 
input into Eq.~(\ref{deter_n}), in the same way as we did originally for 
$i^{*}_{\ell}$, we see that as long as the cities are connected by non-zero 
$c_{jk}$, then they will have no infected individuals present. From 
Eq.~(\ref{first_FP_eqn}) it follows that $s^{*}_k = 1$ for these cities. This 
is the trivial solution where no infection is present anywhere in this cluster 
of connected cities. We will assume that all the cities are connected either 
directly or indirectly, so that $ i^{*}_k = 0, s^{*}_k = 1$ for all $k$.  

Of more interest is what we will call ``the symmetric fixed point''. This
has $s^{*}_k = s^{*}$, a constant, for all $k$. From 
Eq.~(\ref{first_FP_eqn}) one sees that the $i^{*}_k$ are also independent 
of $k$, and we denote them by $i^{*}$. Using Eq.~(\ref{relations_n}), $s^{*}$
and $i^{*}$ are found to satisfy the equations
\begin{eqnarray}
& & s^{*} \left[ \left( \beta + \gamma + \mu \right) -\beta s^{*} \right] = 
\left( \gamma + \mu \right), \nonumber \\
& & \left( \gamma + \mu \right)i^{*} = \mu\left( 1 - s^{*} \right),
\label{one_city}
\end{eqnarray}
which are the fixed point equations for the ordinary `one-city' SIR model
\cite{and_may,KR07}. As is well known these may be solved to give for the 
non-trivial fixed point
\begin{equation}
s^*=\dfrac{\gamma+\mu}{\beta}, \ \ 
i^*=\dfrac{\mu\left[\beta-(\gamma+\mu)\right]}{\beta(\gamma+\mu)}.
\label{FP}
\end{equation}

Due to a remarkable theorem, we can assert that the symmetric solution 
given by Eq.~(\ref{FP}) is the only non-trivial fixed point of the 
deterministic equations (\ref{deter_n}) \cite{guo-etal2008}. This is proved 
by finding a Liapunov function for the $n$-city SIR model. In fact the result 
is more general than we require and was proved for the SEIR model; in 
Appendix~\ref{app:b} we give the explicit form of the Liapunov function for 
the SIR model and a brief outline of the proof following the argument in 
Ref.~\cite{guo-etal2008} for this simpler case. The theorem also tells us that 
the non-trivial fixed point (\ref{FP}) is globally stable. Therefore we can 
now go on to study stochastic fluctuations about this well characterized 
attractor.

\section{Spectrum of the Stochastic Fluctuations}
\label{results}
Based on previous studies of stochastic fluctuations in the SIR model in 
different contexts, we would expect that the fixed point behavior predicted in 
the deterministic limit is replaced by large stochastic oscillations 
\cite{mckane_new2,alonso}. In effect, the noise due to the randomness of the 
processes in the IBM, sustains the natural tendency for cycles to occur, and 
amplifies them through a resonance effect. Since the oscillations are 
stochastic, straightforward averaging will simply wipe out the cyclic 
structure; to understand the nature of the fluctuations we need to Fourier 
transform them and then pick out the dominant frequencies.

So we begin by taking the Fourier transform of the linear stochastic 
differential equation Eq.~(\ref{Langevin}) (generalized to the case of $n$ 
cities) to find
\begin{equation}
\sum^{2n}_{K=1} \left( -i\omega \delta_{JK} - A_{JK} \right) \tilde{z}_{K} 
(\omega) = \tilde{\eta}_{J}(\omega), \ \ J=1,\ldots,2n,
\label{FT_Lang}
\end{equation}
where the $\tilde{f}$ denotes the Fourier transform of the function $f$.
Defining the matrix $-i\omega \delta_{JK} - A_{JK}$ to be $\Phi_{JK}(\omega)$,
the solution to Eq.~(\ref{FT_Lang}) is
\begin{equation}
\tilde{z}_{J}(\omega) = \sum^{2n}_{K=1} \Phi^{-1}_{JK}(\omega) 
\tilde{\eta}_{K}(\omega).
\label{solution}
\end{equation}
  
The power spectrum for fluctuations carrying the index $J$ is defined by 
\begin{equation}
P_{J}(\omega) \equiv \left\langle | \tilde{z}_J(\omega) |^{2} \right\rangle
= \sum^{2n}_{K=1} \sum^{2n}_{L=1} \Phi^{-1}_{JK}(\omega)B_{KL}
\left( \Phi^{\dag} \right)^{-1}_{LJ}(\omega).
\label{PS_defn}
\end{equation}
Since $\Phi = -i\omega I - A$, where $I$ is the $2n \times 2n$ unit matrix, and
since $A$ and $B$ are independent of $\omega$, the structure of $P_{J}(\omega)$
is that of a polynomial of degree $4n-2$ divided by another polynomial of 
degree $4n$. The explicit form of the denominator is 
$|\det \Phi (\omega) |^{2}$.

Oscillations with well-defined frequencies should show up as peaks in the 
power-spectrum. The structure of the power spectrum described above --- with 
the ratio of polynomials of high order potentially giving rise to many 
maxima --- might lead us to suppose that the spectrum of fluctuations would 
have a rather complex structure. In fact numerical simulations indicate that 
only a single peak is present for a large range of parameter values. An 
example is shown in Fig.~3, where typical values for measles 
\cite{BauchEarn,KeelingGrenfellMeasles,and_may} were chosen for the 
epidemiological parameters (we shall keep these values fixed throughout this 
section).


\begin{figure}
\centering
\includegraphics[width=0.5\columnwidth]{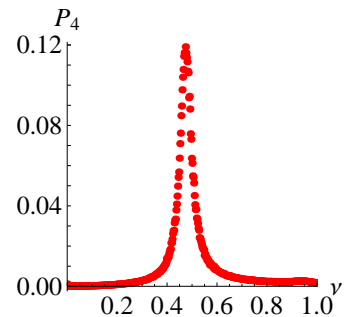}
{\caption{(Color online) Power spectrum for the fluctuations of infectives 
from simulation of a three-city model with equal population sizes plotted as 
a function of the frequency $\nu=\omega/(2\pi)$ 1/y. The  spectrum shown 
corresponds to city 1, the spectra for the other cities are very similar. 
Metapopulation model parameters: $N_1=N_2=N_3=10^6$, $c_{12}=0.06$, and 
$c_{13}=c_{23}=0.02$. Epidemiological parameters: $\gamma=365/13$ 1/y, 
$\mu=1/50$ 1/y, and $\beta= 17 (\gamma+\mu)$.}}
\label{fig3}
\end{figure}


To understand how this comes about, we first note that the number of peaks
in the power spectrum is given by the form of the denominator, 
$|\det \Phi (\omega) |^{2}$; the numerator essentially just shifts the position
of these peaks somewhat. Therefore we can understand the number and nature 
of the peaks by studying the eigenvalues of $\Phi_{JK}$, which are those of
the matrix $A_{JK}$ shifted by $-i\omega$. 

Each pair of complex conjugate eigenvalues of 
$A_{JK}$, $\lambda_c, \lambda_c^*$, will give rise to a factor in 
$|\det \Phi (\omega) |^{2}$ of the form 
$(|\lambda_c|^2 - \omega ^2)^2 + \left[ 2\operatorname{Re}(\lambda_c) \omega \right]^2$, and each real eigenvalue of $A_{JK}$, $\lambda_r$, yields a factor of 
the form $(\lambda_r^2 + \omega ^2)^2$. Peaks in the power spectrum are 
associated with complex eigenvalues $\lambda_c$ of $A_{JK}$ with small real 
parts, and their position is approximately given by 
$\omega \approx \operatorname{Im} (\lambda_c)$. In the trivial case of one 
city, $n=1$, $A_{JK}$ has a pair of complex conjugate eigenvalues 
$\lambda_1^{\pm}$ with  
$\operatorname{Re}(\lambda_1^{\pm})= - \beta \mu /(2(\gamma + \mu))$ and $|\lambda_1^{\pm}| = \sqrt{\mu(\beta -\gamma - \mu)}$ (see Appendix~\ref{app:b}). The 
conditions for a pronounced peak for $\omega $ close to 
$\operatorname{Im}(\lambda_1^{\pm}) \approx |\lambda_1^{\pm}|$ are fulfilled 
because $\mu $, the death-birth rate, is small. This carries over to the 
general $n$ city case since, as shown in Appendix~\ref{app:b}, 
$\lambda_1^{\pm}$ always belong to the set of eigenvalues of $A_{JK}$.  For 
the parameter values of Fig.~3 the numerical values of the common eigenvalue 
pair are $\lambda_1^{\pm}=-0.17\pm i\, 2.99 $, so we expect a peak to be 
located close to $\nu = \operatorname{Im}(\lambda_1^{\pm})/(2\pi) \approx 0.48$
1/y.


\begin{figure}
\centering
\includegraphics[width=0.5\columnwidth]{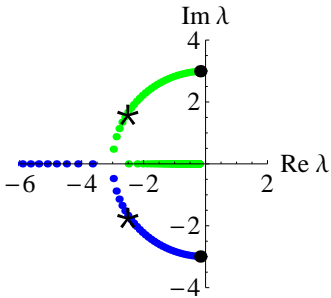}
{\caption{(Color online) An Argand diagram of the eigenvalues for the two-city 
model with $q=N_2/N_1=3/2$ and $c_{12}\in[0,1]$. The large black dots are the 
common eigenvalue pair $\lambda_1^{\pm}$. The sets of smaller dark gray (blue) 
and light gray (green) dots are the remaining eigenvalues $\lambda_2^{\pm}$ computed on a 
uniform grid of values of $c_{12}$ in the interval. The eigenvalues with 
$\operatorname{Re} \lambda_2^{-} <-6$ are not shown in the plot, they are 
found for $c_{12}>0.15$. The asterisks show the eigenvalues for the parameter 
values used in Fig.~5.}}
\label{fig4}
\end{figure}


For large demographic coupling, the $n$ city system will behave as well mixed 
system comprising all the cities and we expect to find in that limit a power 
spectrum similar to the case $n=1$, where each city contributes proportionally 
to its size to the overall spectral density. In the opposite limit, the $n$ 
cities uncouple and we will find for each city the power spectrum of the one 
city case. In order to understand why additional peaks do not show up in 
simulations for intermediate coupling strengths, it is useful to consider 
the case $n=2$, for which the eigenvalues of $A_{JK}$ can be determined 
analytically and depend on a single coupling parameter $c_{12}$ and the ratio 
of the population sizes $q=N_2/N_1$ (see Eq.~(\ref{relations_2}) and 
Appendix~\ref{app:b}). An Argand diagram of the two pairs of eigenvalues, 
$\lambda_1^{\pm}$ and $\lambda_2^{\pm}$, for the two-city model is shown in 
Fig.~4. It can be seen that as the coupling increases, $\lambda_2^{\pm}$ 
follow the circle ${\cal C}$  centered at zero that goes through 
$\lambda _1^{\pm}$, moving away from the imaginary axis. Real and 
imaginary parts become of the same order for very small values of the coupling, 
and so we expect the power spectrum to be always dominated by the peak 
associated with $\lambda_1^{\pm}$ that characterizes the spectrum in the 
uncoupled case. This behavior carries over to the $n$-city case with symmetric 
coupling, for which a complete analysis of the eigenvalues of $A_{JK}$ can 
also be given, see Appendix~\ref{app:b}. In particular, it can be shown that 
apart from the common eigenvalue pair $\lambda _1^{\pm}$ $A_{JK}$, has a single 
$(n-1)$-fold degenerate additional eigenvalue pair that behaves as a function 
of the coupling parameter as described above for $n=2$.

For the coupling parameter that corresponds to the values of $\lambda_2^{\pm}$ 
marked with asterisks in Fig.~4 and for a certain choice of population 
sizes, the infective fluctuations power spectra for the two-city 
model obtained from simulations and from Eq.~(\ref{PS_defn}) are shown in 
Fig.~5. We find a nearly perfect match between the results of numerical 
simulations and the analytical calculations. In agreement with the 
above argument the power spectra of city $1$ and city $2$ are very similar to 
the power spectrum of the one city case, which in turn is very similar to the 
spectrum shown in Fig.~3 for 3 cities with small coupling. In all cases 
the functional form of the spectral density is dominated by the peak 
associated with the common eigenvalue pair $\lambda_1^{\pm}$. As for the 
amplitudes of the power spectra $P_J(\nu)$, their ratio with respect to the 
one city case, $r_J(\nu )$, decreases as the coupling increases. For two 
cities and $q=1$, the power spectra $P_3$ and $P_4$ of city $1$ and city $2$ 
are equal and the relative peak amplitudes  $r_{3,4}(\nu_{\rm max} )$ decrease 
with the coupling strength $c_{12}$ down to $0.5$. For other values of $q$, 
as in Fig.~5, the different peak amplitudes in two cities reflect the symmetry 
$P_3(\nu; c_{12}, q) = P_4(\nu; c_{12}/q, 1/q)$. Depending on $q$, the ratio 
$r_{3,4}(\nu)$ may become even smaller than $0.5$, but due to the symmetry 
that relates $P_3$ and $P_4$, the amplitude of at least one of these peaks is 
always comparable to that of the uncoupled case. More precisely, it is easy 
to check that $1 \leq r_3(\nu; c_{12}, q) + r_4(\nu; c_{12}, q) \leq 2$, where 
the second inequality is satisfied strictly for $c_{12} = 0$ and the lower 
bound corresponds to the large coupling limit $c_{12} = 1$ and to
$\nu = \nu_{\rm max}$ .


\begin{figure}
\centering
\includegraphics[width=\columnwidth]{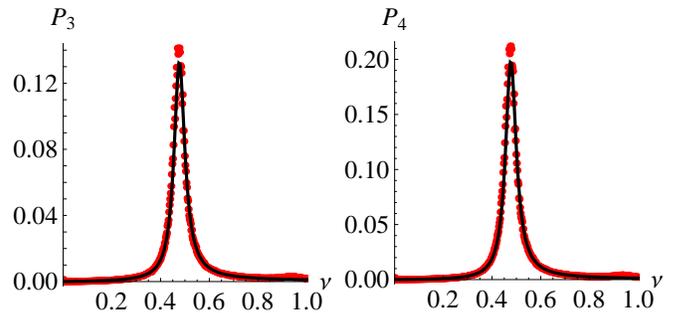}
{\caption{(Color online) Power spectra for the fluctuations of infectives 
from simulation of the two-city model [(red) dots] and analytic calculation 
(black solid curve) plotted as a function of the frequency $\nu=\omega/(2\pi)$ 
1/y. The population sizes were chosen to be $N_1=10^6$ and $N_2=1.5\times10^6$ 
so that their ratio is $3/2$. The coupling coefficient $c_{12}=0.1$. The 
location of the eigenvalues for this choice of parameters is indicated in 
Fig.~4 by asterisks and large dots.}}
\label{fig5}
\end{figure}



\begin{figure}
\centering
\includegraphics[width=\columnwidth]{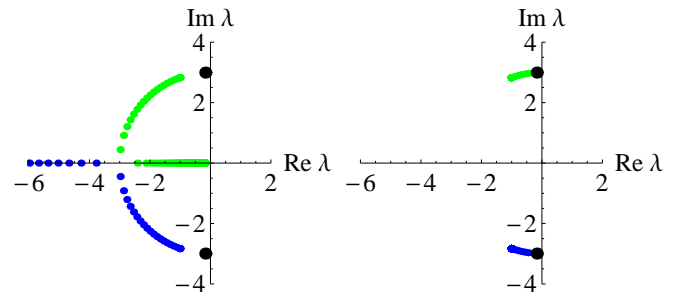}
{\caption{(Color online) An Argand diagram of the eigenvalues for a three-city 
model with equal population sizes, $c_{12}\in[0,0.98]$ and the other parameters
as in Fig.~3. The large black dots are the common eigenvalue pair 
$\lambda_1^{\pm}$. The sets of smaller dark gray (blue) and light gray (green) dots are 
the remaining eigenvalues $\lambda_2^{\pm}$ (left panel) and $\lambda_3^{\pm}$ 
(right panel) computed on a uniform grid of values of $c_{12}$ in the interval.
The eigenvalues with $\operatorname{Re} \lambda_2^{-} <-6$ are not shown in the
plot, they are found for $c_{12}>0.12$.}}
\label{fig6}
\end{figure}


The general case of three cities with no symmetry can also in principle be 
treated analytically because finding the eigenvalues of $A_{JK}$ reduces to 
finding the roots of a fourth order polynomial. However, the problem now 
depends on 3 independent coupling parameters and 2 parameters for city sizes 
and closed form expressions are too lengthy to be useful. An approximate, 
concise description of the behavior of the eigenvalues of $A_{JK}$ can be 
given in terms of only two parameters that measure coupling strength and 
coupling asymmetry, see Appendix~\ref{app:b}. In this approximation, we 
assume that all the $c_{jk}$, $j \neq k $, are of order $\sqrt{\mu}$ and 
treat $\mu$ as the small parameter of the system. Simple expressions for the 
real parts and the absolute values of the additional eigenvalue pairs 
$\lambda_2^{\pm}$, $\lambda_3^{\pm}$ of $A_{JK}$  up to terms of order $\mu$ can 
be derived [see Eqs.~(\ref{approxeigenre}) and (\ref{modulus})]. These show 
that, in this approximation, both eigenvalue pairs behave as described for 
the symmetric case. As the coupling increases, both eigenvalue pairs follow 
the circle ${\cal C}$  centered at zero that goes through $\lambda _1^{\pm}$, 
moving away from the imaginary axis. The real and imaginary parts become of 
the same order within the scope of the approximation. 
Equation (\ref{approxeigenre}) also shows how the asymmetry lifts the 
degeneracy of the two pairs $\lambda_2^{\pm}$, $\lambda_3^{\pm}$. As the 
coupling increases, the two eigenvalue pairs move along the circle ${\cal C}$ 
at different speeds. We have checked that Eqs.~(\ref{approxeigenre}) and 
(\ref{modulus}) give a good approximation to the exact results in the regime 
when the eigenvalues are complex.

The same behavior is illustrated in Fig.~6, where a plot of the exact 
solutions for $\lambda_{2,3}^{\pm}$ is shown for parameter values that 
correspond to taking those of Fig.~3 and allowing one of the coupling 
coefficients to span the whole admissible range. One of the eigenvalues is 
shown only up to $c_{12}=0.12$, where its real part becomes smaller than $-6$. 


\begin{figure}
\centering
\includegraphics[width=0.5\columnwidth]{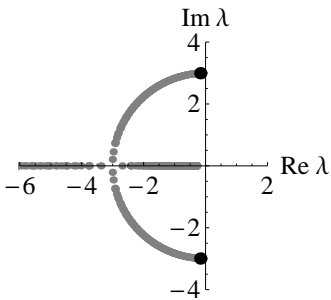}
{\caption{An Argand diagram of the eigenvalues for a four-city model with the 
coupling strength $x\in[0,0.52]$. The large black dots are the common 
eigenvalue pair $\lambda_1^{\pm}$. The remaining eigenvalues 
$\lambda_{2,3,4}^{\pm}$ computed on a uniform grid of values of $x$ in the 
interval are shown as sets of smaller gray dots. As in the previous figures 
we only show eigenvalues whose real part is larger than $-6$. Metapopulation 
model parameters: $N_2/N_1:N_3/N_1: N_4/N_1=2:3:4$, ${\hat c}_{12}=1/\sqrt{\mu}=2{\hat c}_{13}=5 {\hat c}_{14}/2=5 {\hat c}_{23}/2=3{\hat c}_{24}=4{\hat c}_{34}$.}}
\label{fig7}
\end{figure}


In Fig.~7 we show numerical results for the behavior of the eigenvalues 
of $A_{JK}$ in the case of 4 cities with different population sizes
and a certain choice of the coupling coefficients $c_{jk}$, $j,k=1,2,3,4$. We 
make use of the following notation for the diagonal and off diagonal coupling 
coefficients (see Appendix ~\ref{app:b}): 
$c_{jj}=1-{\hat c}_{jj} \ x \ \sqrt{\mu}$ and 
$c_{jk}={\hat c}_{jk}\ x \ \sqrt{\mu }$, respectively. We then calculate 
the set of three non-trivial eigenvalue pairs as the coupling strength $x$ 
varies in a suitable interval, keeping the ${\hat c}_{jk}$ fixed. These results 
suggest that the behavior of the eigenvalues of $A_{JK}$ is essentially given 
by the description of the symmetric case, and that more general couplings 
break the degeneracy as in the case $n=3$, with no effects in the contributions 
to the peaks in the power spectrum. 

\section{Discussion and conclusions}  
\label{conclude}
In this paper we have extended the analysis of a metapopulation model 
of epidemics into the stochastic domain. Frequently epidemic models involving
a spatial component, such as the interaction between several cities, are 
studied purely deterministically \cite{mech_approach,multi_city} or through 
computer simulations \cite{lloyd_may,STdG07,grenfellphysA2010}. We have 
demonstrated how a stochastic metapopulation model can be studied analytically 
by using a relatively straightforward extension of the methodology which was 
used to study a well-mixed population in a single city.

We adopted a simple specification of residents and commuters in order to set
up the model. However, the coefficients which appear in the dynamical equations
are generic and would appear in the same form if residents and commuters were
included in a different way. It is evident that there are many ways of 
characterizing the interchange of individuals between cities which will result
in the same model; only the identification of the coefficients with the
underlying structure will be different.

The deterministic form of the model predicts that the system will reach a 
stable fixed point where the proportion of infected, susceptible and recovered
individuals is the same in every city. The stochastic version of the model  
also predicts a clean simple result: that the large sustained oscillations
which replace the deterministic predictions of constant behavior, have a
single frequency which is the same for every city. Moreover, for small, large 
and intermediate coupling between the cities, the form of the power spectrum 
of these fluctuations is closely approximated by the power spectrum of the 
single city system. 

It is remarkable that such a simple result occurs in what is a quite complicated
stochastic nonlinear metapopulation model. We hope to explore the range of
validity of this result and its robustness to the addition of new features to
the model in the future. In any case, we believe that the work presented here 
will give a firm foundation to possible future work, including comparisons 
with the data available on childhood diseases.

\begin{acknowledgments}
Financial support from the Portuguese Foundation for Science and Technology 
(FCT) under Contract No. POCTI/ISFL/2/261 is gratefully acknowledged. G.R. was
also supported by FCT under Grant No. SFRH/BPD/69137/2010.
\end{acknowledgments}

\appendix

\section{System-size expansion}
\label{app:a}
Here we give some of the key steps in the application of the system-size
expansion to the model explored in this paper. The method has been extensively
discussed in the literature \cite{van_kampen,alonso,STdG07,PhysRevE.79.041922,PhysRevE.80.051915,seasonalRN,parisi,epidforcingmckane,stagedmodelAndrew}, and 
so we confine ourselves to a brief outline and to displaying the most 
important intermediate results in the derivation. We will assume that we are 
carrying out the calculation for the $n$-city case discussed in 
Section~\ref{n_city}; the corresponding results for Section~\ref{two_city} can 
be obtained simply by setting $n=2$.

The first point to mention is that there are apparently $n$ expansion 
parameters: $\{N_1,\ldots,N_n\}$. The method is valid if they are all large
and of the same order. More formally we can take, for instance, $N_{1}\equiv N$
as the expansion parameter and express all the other $N_j$ in terms of it:
$N_{j}=Nq_j$, where the $q_{j}=N_{j}/N$, $j=2,\ldots,n$ are of order one. In
practice the method seems to work well when the $q_j$ are significantly 
different from one, but this has to be checked a posteriori, for instance by
comparing the analytic results with those obtained using computer simulations.
In what follows we will not introduce the $q_j$ explicitly; we will simply 
take all the $N_j$'s to be of the same order in the expansion.

The van Kampen ansatz Eq.~(\ref{ansatz}) replaces the discrete stochastic 
variables $\sigma$ by the continuous stochastic variables $\mathbf{z}$
and so we write the transformed probability distribution $P(\sigma,t)$ as
$\Pi(\mathbf{z},t)$. Since this transformation is time-dependent, substituting
the ansatz into $dP/dt$ on the left-hand side of Eq.~(\ref{master}) 
gives \cite{van_kampen}
\begin{eqnarray}
\frac{dP(\sigma,t)}{dt} &=& \frac{\partial \Pi(\mathbf{z},t)}
{\partial t} - \sum^{n}_{j=1} \sqrt{N_j}\, 
\frac{\partial \Pi(\mathbf{z},t)}{\partial x_j} \frac{ds_j}{dt} \nonumber \\
&-& \sum^{n}_{j=1} \sqrt{N_j}\, 
\frac{\partial \Pi(\mathbf{z},t)}{\partial y_j} \frac{di_j}{dt}\,.
\label{LHS}
\end{eqnarray}

The right-hand side of the master equation (\ref{master}) can be put into a
form from which it is simple to apply the expansion procedure. To do this 
one introduces step-operators \cite{van_kampen} defined by
\begin{align}
\epsilon^{\pm 1}_{S_j}&f(S_1,\ldots,S_j,\ldots,S_n,I_1,\ldots,I_n)
\nonumber \\
 = &f(S_1,\ldots,S_{j}\pm 1,\ldots,S_n,I_1,\ldots,I_n), \nonumber \\
 \epsilon^{\pm 1}_{I_j}&f(S_1,\ldots,S_n,I_1,\ldots,I_j,\ldots,I_n) 
\nonumber \\
 = &f(S_1,\ldots,S_n,I_1,\ldots,I_{j}\pm 1,\ldots,I_n),
\label{step_ops}
\end{align}
for a general function $f$ and where $j=1,\ldots,n$. Using these operators
the master equation (\ref{master}) may be written as
\begin{align}
& \dfrac{dP(\sigma,t)}{dt} = 
\sum_{j=1}^n \left[ \left(\epsilon_{I_j}-1\right)T_j 
+\left(\dfrac{\epsilon_{I_j}}{\epsilon_{S_j}} -1\right)T_{n+j} \right.\nonumber\\
& + \left. \left(\dfrac{1}{\epsilon_{S_j}} -1\right)T_{2n+j}
+ \left(\dfrac{\epsilon_{S_j}}{\epsilon_{I_j}} -1\right)T_{3n+j} \right]
P(\sigma,t).
\label{master_app}
\end{align}

Within the system-size expansion these operators have a simple structure:
\begin{equation}
\epsilon_{S_j} = \sum\limits_{p=0}^{\infty}\dfrac{N_j^{-p/2}}{p!}
\dfrac{\partial^p}{\partial x_{j}^p}, \ \ 
\epsilon_{I_j} = \sum\limits_{p=0}^{\infty}\dfrac{N_j^{-p/2}}{p!}
\dfrac{\partial^p}{\partial y_{j}^p},
\label{step_ops_expand}
\end{equation}
and so all the terms of the right-hand side of Eq.~(\ref{master_app}) may 
be straightforwardly expanded. Comparing these with the left-hand side in
Eq.~(\ref{LHS}) the leading order ($\sim \sqrt{N_j}$) yields the deterministic
equations given by Eq.~(\ref{deter_n}). The next-to-leading order (which is of
order one) gives a Fokker-Planck equation:
\begin{equation}
\frac{\partial \Pi}{\partial t} = 
- \sum^{2n}_{J,K=1} \frac{\partial }{\partial z_J}\left[ A_{JK} z_K \Pi \right] 
+ \frac{1}{2} \sum^{2n}_{J,K=1} B_{JK}
\frac{\partial^2 \Pi}{\partial z_J \partial z_K}.
\label{FPE}
\end{equation}
The $2n \times 2n$ matrices $A$ and $B$ which appear in this equation have the
following form. Writing $A$ in blocks of four $n \times n$ submatrices:
\begin{equation}
A =\left[\begin{array}{c|c} 
A^{(1)} & A^{(2)} \\ \hline 
A^{(3)} & A^{(4)} 
 \end{array}\right],
\label{block}
\end{equation}
the elements of these submatrices are
\begin{eqnarray}
A^{(1)}_{jk}&=&-\mu\delta_{jk}-\beta\delta_{jk}\sum_{\ell=1}^n c_{j\ell}i_{\ell},
\nonumber \\
A^{(2)}_{jk}&=&-\beta\left(\dfrac{N_j}{N_k}\right)^{1/2}s_jc_{jk},
\nonumber \\
A^{(3)}_{jk}&=&\beta\delta_{jk}\sum_{\ell=1}^n c_{j\ell}i_{\ell},
\nonumber \\
A^{(4)}_{jk}&=&-(\mu+\gamma)\delta_{jk}+
\beta\left(\dfrac{N_j}{N_k}\right)^{1/2}s_jc_{jk}.
\label{A_entries}
\end{eqnarray}
Writing $B$ in a similar way to $A$ in Eq.~(\ref{block}), the elements of 
the submatrices are
\begin{eqnarray}
B^{(1)}_{jk} &=& \mu\delta_{jk}\left( 1 - s_j \right) + \beta\delta_{jk} 
\sum_{\ell=1}^n s_j c_{j\ell} i_{\ell}, \nonumber \\
B^{(2)}_{jk}&=&B^{(3)}_{jk}= -\mu\delta_{jk} i_{j} -\beta\delta_{jk} 
\sum_{\ell = 1}^n s_j c_{j\ell} i_{\ell}, \nonumber \\
B^{(4)}_{jk}&=& \left( \gamma + \mu \right)\delta_{jk} i_j + \beta\delta_{jk}
\sum_{\ell =1}^n s_j c_{j\ell} i_{\ell}.
\label{B_entries}
\end{eqnarray}
From Eqs.~(\ref{A_entries}) and (\ref{B_entries}) it is clear that the matrices
$A_{jk}$ and $B_{jk}$ depend on the solutions of the deterministic equations 
given in Eq.~(\ref{deter_n}). However, since we will be interested only in 
fluctuations about the stationary state, these matrices are evaluated at the
fixed point. Since the unique stable fixed point is the symmetric one, the
same for all cities, the entries (\ref{A_entries}) and (\ref{B_entries}) are
given by:
\begin{align}
A^{*(1)}_{jk} &= - \left[ \mu +\beta i^{*} \right]\delta_{jk}, \ \ 
A^{*(2)}_{jk} = -\beta\left(\dfrac{N_j}{N_k}\right)^{1/2}s^{*}c_{jk},
\nonumber \\
A^{*(3)}_{jk} &= \beta i^{*}\delta_{jk}, \ \
A^{*(4)}_{jk} = 
\beta\left(\dfrac{N_j}{N_k}\right)^{1/2}s^{*}c_{jk}-(\mu+\gamma)\delta_{jk}, 
\nonumber \\
\label{A_entries*}
\end{align}
and
\begin{align}
& B^{*(1)}_{jk} = 2\mu \left( 1 - s^{*} \right)\delta_{jk}, \ \ 
B^{*(4)}_{jk} = 2\left( \gamma + \mu \right) i^{*}\delta_{jk}, \nonumber \\
& B^{*(2)}_{jk}=B^{*(3)}_{jk} = - i^{*} \left[ \mu +\beta s^{*}\right]\delta_{jk},
\label{B_entries*}
\end{align}
where we have used the fixed-point equation (\ref{one_city}) to simplify some
of the entries in Eq.~(\ref{B_entries*}). 

Finally, the Fokker-Planck equation (\ref{FPE}) is equivalent to the 
stochastic differential equation (\ref{Langevin}). We will work with the latter,
since we wish to use Fourier analysis to analyze the nature of the fluctuations,
and since Eq.~(\ref{Langevin}) is linear, it can easily be Fourier transformed, 
as discussed in detail in Section~\ref{results}.

\section{Some results for the $n$-city case}
\label{app:b}
In this Appendix we give some of the derivations for the $n$-city case discussed
in Section~\ref{n_city} and Section~\ref{results} which are too long and 
cumbersome to be given in the main text.

\subsection{The coefficients $c_{jk}$}
\label{cs_n}
The coefficients $c_{jk}$ appearing in Eq.~(\ref{infection_n}) may be read off
from the four types of term (i)-(iv) given in Section~\ref{n_city}:
\begin{eqnarray}
c_{jj} &=& \frac{\left( 1 - f_j \right)^2 N_j}{\left[ (1-f_j)N_j + 
\sum_{m\neq j} f_{jm}N_m \right]} \nonumber \\
&+& \sum_{\ell \neq j} \frac{f_{\ell j}^2 N_j}{\left[ (1-f_{\ell})N_{\ell} + 
\sum_{m\neq \ell} f_{\ell m}N_m \right]},
\label{c_jj}
\end{eqnarray}
for $j=1,\ldots,n$ and
\begin{eqnarray}
c_{jk} &=& \frac{\left( 1 - f_j \right)f_{jk} N_k}{\left[ (1-f_j)N_j + 
\sum_{m\neq j} f_{jm}N_m \right]} \nonumber \\
&+& \frac{f_{kj}\left( 1 - f_k \right) N_k}{\left[ (1-f_k)N_k + 
\sum_{m\neq k}f_{km}N_m \right]} \nonumber \\
&+& \sum_{\ell\neq j,k} \frac{f_{\ell j} f_{\ell k} N_k}{\left[ (1-f_{\ell})N_{\ell} 
+ \sum_{m\neq \ell} f_{\ell m}N_{m} \right]},
\label{c_jk}
\end{eqnarray}
for $j,k=1,\ldots,n$ and $j \neq k$.

To prove the first relation given in Eq.~(\ref{relations_n}), consider the
sum $c_{jj} + \sum_{k\neq j}c_{jk}$. The first term in Eq.~(\ref{c_jj}) 
combines with the first term in Eq.~(\ref{c_jk}) to give $(1-f_j)$. The 
last term in Eq.~(\ref{c_jj}) combines with the last term in Eq.~(\ref{c_jk}) 
to give
\begin{eqnarray}
& & \sum_{\ell \neq j} \frac{f_{\ell j} \left[ f_{\ell j} N_j + 
\sum_{k\neq j,\ell} f_{\ell k} N_k \right]}{\left[ (1-f_{\ell})N_{\ell} + 
\sum_{m\neq \ell} f_{\ell m}N_m \right]}  \nonumber \\
& & = \sum_{\ell \neq j} \frac{f_{\ell j} 
\left[ \sum_{k \neq \ell} f_{\ell k} N_k \right]}{\left[ (1-f_{\ell})N_{\ell} + 
\sum_{m\neq \ell} f_{\ell m}N_m \right]}  \nonumber \\
& & = \sum_{k \neq j} \frac{f_{k j} \left[ \sum_{m \neq k} f_{km}N_m \right] }
{\left[ (1-f_{k})N_{k} + \sum_{m\neq k} f_{k m}N_m \right]},
\label{manipulate}
\end{eqnarray}
where in the last line we have performed a relabeling. Combining the middle 
term of Eq.~(\ref{c_jk}) with the result in Eq.~(\ref{manipulate}) gives
\begin{equation}
\sum_{k \neq j} \frac{f_{kj} \left[ \left( 1 - f_k \right) N_k + \sum_{m \neq k} 
f_{km}N_m \right] }{\left[ (1-f_k)N_k + 
\sum_{m\neq k}f_{km}N_m \right]} = f_{j},
\label{penulti}
\end{equation}
using Eq.~(\ref{f_j}). Adding this to the result $(1-f_j)$ found earlier 
proves the result $c_{jj} + \sum_{k\neq j}c_{jk} = 1$.

We also note from Eq.~(\ref{c_jk}) that $c_{jk}/N_k$ is symmetric under the 
interchange of $j$ and $k$ Therefore
\begin{equation}
\frac{c_{jk}}{N_k} = \frac{c_{kj}}{N_j},
\label{second_proof}
\end{equation}
which is the second relation in Eq.~(\ref{relations_n}).

\subsection{Uniqueness and stability of the fixed point}
\label{unique}
In Section~\ref{n_city} we asserted that the deterministic equations 
(\ref{deter_n}) have a unique non-trivial fixed point, which was globally 
stable. Here we prove this by giving a Liapunov function for the dynamical 
system in the invariant region $R=\{ (s_1,\ldots,s_n,i_1,\ldots,i_n) : 
0\leq s_j \leq 1, 0\leq i_j \leq 1, s_j + i_j \leq 1, j=1, ...,n \}$
where the system is defined. This is a modification of the function given in 
Ref.~\cite{guo-etal2008} for the SEIR model. The proof assumes that the matrix 
of the coupling coefficients $c_{jk}$ is irreducible, which means that any two 
cities have a direct or indirect interaction. Otherwise the proof breaks down 
because the $n$ cities may be split into non-interacting subsets, and several 
equilibria may be found by combining disease extinction in some subsets with 
non-trivial equilibrium in other subsets. 

Let $\beta_{jk} \equiv \beta c_{jk} s^*_j i^*_k$, where 
$(s^*_1,\ldots,s^*_n,i^*_1,\ldots,i^*_n)$ is a fixed point of 
Eq.~(\ref{deter_n}), and denote by $M$ the matrix defined by 
$M_{kj}= \beta_{jk}, j\neq k$, and $\sum_{k=1}^n  M_{kj} =0, j=1, ...,n$. It can 
be shown (\cite{guo-etal2008}, Lemma 2.1) that the solution space of the 
linear equation $M v = 0$ is spanned by a single vector $(v_1,\ldots,v_n)$, 
$v_j > 0, j=1,\ldots,n$. Let $L(s_1,\ldots,s_n,i_1,\ldots,i_n)$ be defined as
\begin{equation}
L = \sum_{j=1}^n v_j ( s_j - s_j^* \log s_j + i_j - i_j^* \log i_j).
\nonumber
\end{equation}
$L$ has a global minimum in $R$ at the fixed point. Functions of this form 
have been used in the literature as Liapunov functions for fixed points of 
ecological and epidemiological models, whose variables take only positive 
values \cite{guo-etal2008}. Differentiating $L$ along the solutions of 
Eq.~(\ref{deter_n}), and following the proof of Theorem 1.1 in 
Ref.~\cite{guo-etal2008}, we obtain
\begin{equation}
\dot L \leq  \sum_{j,k=1}^n v_j M_{kj} \left (
2 - \frac{s_j^*}{s_j} - \frac{s_j}{s_j^*} 
\frac{i_k}{i_k^*} \frac{i_j^*}{i_j} \right ).
\label{boundliap}
\end{equation}
The properties of the coefficients $v_j$ in the definition of $L$ play a 
crucial role in the derivation of the second term in this inequality. Use has 
been made of the identity
\begin{equation}
\sum_{j=1}^n v_j \sum_{k=1}^n \beta c_{jk} s^*_j i_k = \sum_{j=1}^n v_j 
(\gamma + \mu ) i_j,
\end{equation}
which in turn uses the fact that $M v = 0$ can be written as
\begin{equation}
\sum_{j=1}^n  \beta c_{kj} s^*_k i^*_j v_k = \sum_{j=1}^n  
\beta c_{jk} s^*_j i^*_k v_j, \ , k=1,\ldots,n.
\end{equation}
Following Ref.~\cite{guo-etal2008}, it can then be shown that the right-hand 
side of Eq~(\ref{boundliap}) is strictly negative except at 
$(s^*_1,\ldots,s^*_n,i^*_1,\ldots,i^*_n)$. Therefore, $L$ is a Liapunov 
function for this fixed point in $R$, and the fixed point is unique and 
globally stable. Note that the result also holds when the disease 
transmissibility $\beta $, the recovery rate $\gamma $ and the birth-death 
rate $\mu $ are different in different cities, in which case the non-trivial 
equilibrium is in general not symmetric.

\subsection{Nature of the eigenvalues of the matrix $A$}
\label{eigen_of_A}
In this subsection we will give some results on the eigenvalues of $A$ which
are required for the discussion in Section~\ref{results}. 

We first recall that $A$ is closely related to the stability matrix of the
deterministic equations (\ref{deter_n}). In fact, in most applications of 
the system-size expansion they are equal. In our case because we have $n$ 
expansion parameters $\sqrt{N}_j$, they are not equal, but closely related. 
A simple calculation of the Jacobian, $J$, from Eq.~(\ref{deter_n}), shows that
\begin{equation}
J= S^{-1} A S, \ \ {\rm where\ } 
S = {\rm diag}\left(\sqrt{N}_1,\ldots,\sqrt{N}_n\right).
\label{similarity}
\end{equation}
The effect of the transformation is simply that one can obtain $J$ from $A$
by omitting the terms $(N_j/N_k)^{1/2}$ in $A^{(2)}_{jk}$ and $A^{(4)}_{jk}$
in Eq.~(\ref{A_entries}) or in $A^{*(2)}_{jk}$ and $A^{*(4)}_{jk}$ in  
Eq.~(\ref{A_entries*}). This is useful, since it follows from the 
similarity transformation (\ref{similarity}) that the eigenvalues of $A$ are
also the eigenvalues of $J$. So we may study the simpler problem of finding
the eigenvalues of the Jacobian at the symmetric fixed point (\ref{FP}). 

For orientation, let us explicitly calculate the characteristic polynomial of
the Jacobian for the cases of one city and two cities. These are 
\begin{equation}
n=1: \ \  \ R_{1}(\lambda)=Q^{-1}(d_{2}\lambda^2+d_1\lambda+d_0),
\end{equation}
where
\begin{eqnarray}
Q&=&\gamma+\mu,\ \ d_{2}=\gamma+\mu,\ \ d_{1} = \beta\mu,\nonumber\\
d_{0}&=&\mu(\gamma+\mu)\left[\beta-(\gamma+\mu)\right],
\label{one_city_poly}
\end{eqnarray}
and
\begin{equation}
n=2: \ \ \ R_{2}(\lambda)=Q^{-2}(d_{2}\lambda^2+d_1\lambda+d_0)
(g_{2}\lambda^2+g_1\lambda+g_0).\nonumber
\end{equation}
Thus,
\begin{equation}
R_{2}(\lambda)=R_{1}(\lambda)Q^{-1}(g_{2}\lambda^2+g_1\lambda+g_0),
\end{equation}
where
\begin{eqnarray}
g_{2}&=&\gamma+\mu,\ \ g_{1}=\beta\mu+(c_{12}+c_{21})(\gamma+\mu)^2,\nonumber\\
g_{0}&=&\mu(\gamma+\mu)\left[\beta-(1-c_{12}-c_{21})(\gamma+\mu)\right].
\label{two_cities}
\end{eqnarray}
We see that the factor $R_{1}(\lambda)$ is common, which suggests that the
pair of eigenvalues found in the one city case might always be present in
the $n$ city case. This is easily proved by considering the vector
$\mathbf{v}=(v_1,\ldots,v_n,v_{n+1},\ldots,v_{2n})^T$ with components satisfying
$v_{i}=v$ and $v_{i+n}=v'$ for $i=1,\ldots,n$. Then the eigenvector equation 
$J^*{\bf v} = \lambda {\bf v}$ reduces to that for one city as required. 

A similar method can be used to find the characteristic polynomial for 
$n\geq 3$ cities with equal population sizes, where the couplings are equal, 
that is,
\begin{equation}
c_{jk}=\begin{cases}
1-(n-1)c, & j=k,\\
c, & j\neq k,
\end{cases}
\end{equation}
where $j,k=1,...,n$.
We now take the components of the vector to be 
$v_1=-v_2=v$, $v_{n+1}=-v_{n+2}=v'$, and $v_i=v_{i+n}=0$ for $i=3,\ldots,n$. The 
eigenvector equation $J^*{\bf v} = \lambda {\bf v}$ now reduces to
\begin{eqnarray}
-\left(\dfrac{\beta\mu}{\gamma+\mu}+\lambda\right)v-(1-nc)(\gamma+\mu)v'
&=&0,\nonumber\\
\left(\dfrac{\beta\mu}{\gamma+\mu}-\mu\right)v-\left[nc(\gamma+\mu)+
\lambda\right]v'&=&0.
\end{eqnarray}
Therefore, both solutions of 
\begin{equation}
Q^{-1}\left(h_2\lambda^2+h_{1n}\lambda+h_{0n}\right)=0,
\label{B.16}
\end{equation}
where
\begin{eqnarray}
h_{2}&=&\gamma+\mu,\ \ h_{1n}=\beta\mu+nc(\gamma+\mu)^2,\nonumber\\
h_{0n}&=&\mu(\gamma+\mu)\left[\beta-(1-nc)(\gamma+\mu)\right],
\end{eqnarray}
are eigenvalues of $J^*$. This procedure can be repeated for $n-1$ independent 
vectors with only four non-zero components and the same symmetry as ${\bf v}$.
Therefore, the characteristic polynomial of $J^*$, $R_{n}(\lambda)$, factorizes
as
\begin{equation}
R_{n}(\lambda)=R_1(\lambda)
\left[Q^{-1}\left(h_2\lambda^2+h_{1n}\lambda+h_{0n}\right)\right]^{n-1}.
\end{equation}

Finally let us consider 3 cities with arbitrary coupling and study the 
eigenvalues of $J^*$ in the limit when the off-diagonal coefficients $c_{jk}$ 
are small and of the same order. It will become clear that the coupling range 
to explore corresponds to $c_{jk}$ of the order of $\sqrt{\mu }$ and it is 
convenient to introduce the notation 
\begin{equation}
c_{jk}(x)=\begin{cases}
1 -  {\hat c}_{jj}\ x  \ \sqrt{\mu }, & j=k,\\
  
{\hat c}_{jk}\ x \ \sqrt{\mu }, & j\neq k,
\end{cases}
\label{family}
\end{equation}
where $j,k=1,2,3$ and $x$ is a positive parameter. Eq.~(\ref{family})
represents, for each choice of ${\hat c}_{jk} $, a family of systems with all 
the off-diagonal coefficients $c_{jk}$ of the same order, that reaches the zero 
coupling limit for $x=0$. The quantity $x$ measures the distance to zero 
coupling along each particular family, scaled by $\sqrt{\mu }$. Taking into 
account the properties of the matrix $c_{jk}$, given by 
Eq.~(\ref{relations_n}), the characteristic polynomial of $J^*$ is a polynomial
of degree six that can be expressed in terms of this distance 
$x \ \sqrt{\mu }$ and of three other independent parameters. We choose these 
to be ${\hat c}_{jj}$, $j=1,2,3$. We know that this characteristic polynomial 
factorizes as 
$R_1(\lambda)(\lambda^4 + p_3 \lambda^3 + p_2 \lambda^2 + p_1 \lambda + p_0)$, 
where $p_3$, $p_2$, $p_1$ and $p_0$ are some coefficients. The roots of 
$R_1(\lambda)$ are the pair of eigenvalues $\lambda _1^{\pm}$ shared by all 
the characteristic equations of this family of systems. The polynomial of 
degree four can be easily found by direct computation. For equal city sizes 
we obtain for the coefficients 
\begin{eqnarray}
p_3 & = & \gamma \sigma \ x \ \sqrt{\mu } + {\cal O}(\mu),\nonumber\\
p_2 & = & 2 (\beta - \gamma) \ \mu  - 3/4 \ \gamma ^2  {\hat p}_2 \ x^2 \ \mu  
+ {\cal O}(\mu ^{3/2}),\nonumber\\
p_1 & = & \gamma (\beta - \gamma) \sigma \ x \  \mu ^{3/2} + 
{\cal O}(\mu ^2),\nonumber\\
p_0 & = & (\beta - \gamma )^2 \ \mu ^2 + {\cal O}(\mu^{5/2}),
\label{coeffsmu}
\end{eqnarray}
where
\begin{eqnarray}
\sigma & = & {\hat c}_{11} + {\hat c}_{22} + {\hat c}_{33}, \nonumber \\  
{\hat p}_2 & = & {\hat c}_{11}^2 +{\hat c}_{22}^2+
{\hat c}_{33}^2-2({\hat c}_{11}{\hat c}_{22}+{\hat c}_{11}{\hat c}_{33}+
{\hat c}_{22}{\hat c}_{33}). \nonumber \\
& &
\end{eqnarray}
Keeping only the leading order terms in each of the coefficients given by 
Eq.~(\ref{coeffsmu}) we find a simple approximate expressions for the two 
additional eigenvalue pairs $\lambda _2^{\pm}$, $\lambda _3^{\pm}$. In 
particular, we find
\begin{eqnarray}
\operatorname{Re}(\lambda _2^{\pm}) & = & - \frac{\gamma }{4} (\sigma + k)\ 
x\ \sqrt{\mu } + {\cal O}(\mu),\nonumber\\
\operatorname{Re}(\lambda _3^{\pm})& = & - \frac{\gamma }{4}(\sigma - k)\ 
x \ \sqrt{\mu } + {\cal O}(\mu),
\label{approxeigenre}
\end{eqnarray}
where
\begin{equation}
k^2  = 4 ({\hat c}_{11}^2 + {\hat c}_{22}^2 + {\hat c}_{33}^2 - 
{\hat c}_{11}{\hat c}_{22} - {\hat c}_{11}{\hat c}_{33}- 
{\hat c}_{22}{\hat c}_{33}).
\label{keq}
\end{equation}
Assuming without loss of generality that ${\hat c}_{33} \geq {\hat c}_{22} \geq 
{\hat c}_{11}$, $k^2 $ is positive and so $k$ is real. Note that $k=0$ 
in the symmetric case, and in that case (\ref{approxeigenre}) coincide in the 
same order of approximation with the roots of Eq.~(\ref{B.16}) for $n=3$. 
The quantities $\sigma\,x\,\sqrt{\mu }$ and $k\,x\,\sqrt{\mu }$ that 
determine, in this approximation, the real parts of the two non-trivial 
eigenvalue pairs can be interpreted as the overall coupling strength and the 
coupling asymmetry for a system of family (\ref{family}). We also find for 
the absolute value of the eigenvalues
\begin{equation}
\vert \lambda _{2,3}^{\pm} \vert = \sqrt{\beta - \gamma} \sqrt{\mu} + 
{\cal O}(\mu),
\label{modulus}
\end{equation}
which shows that, for all families of the form Eq.~(\ref{family}), the 
eigenvalues $\lambda _{2,3}^{\pm}$ of $J^*$ move close to the circle ${\cal C}$ 
in the complex plane centered at zero that goes through $\lambda _1^{\pm}$. 
For arbitrary city sizes, the same calculation can be carried out to find 
that Eq.~(\ref{approxeigenre}) and Eq.~(\ref{modulus}) still hold, with 
Eq.~(\ref{keq}) replaced by
\begin{equation}
k^2  = \sigma^2 + \frac{1 + q_{21} + q_{31}}{q_{21} q_{31}} {\tilde k}^2,
\label{kneq}
\end{equation}
where $q_{jk} = N_j/N_k $ and
\begin{equation}
{\tilde k}^2 = {\hat c}_{11}^2 + ({\hat c}_{22} q_{21} - {\hat c}_{33}  q_{31})^2 
- 2 {\hat c}_{11} ({\hat c}_{22} q_{21} + {\hat c}_{33} q_{31}).
\end{equation}

The behavior of the two non-trivial eigenvalue pairs along a family 
(\ref{family}) can be described, in this approximation, in terms of the two 
parameters $\sigma $ and $k$ that characterize the family and of the scaled 
distance $x$. As $x$ increases away from zero, both eigenvalue pairs move along 
${\cal C}$ with speeds whose ratio is given by $(\sigma  + k )/(\sigma - k)$. 
The parameter $k$ that measures the asymmetry of the coupling causes the 
splitting of the two pairs with respect to the degenerate, symmetric case. The 
first pair to reach the real axis does so for 
$x = 4 \sqrt{\beta - \gamma}/(\gamma (\sigma + k))$, which lies within the 
scope of the approximation. From then on the two real eigenvalues keep 
changing with $x$ in such a way that the square root of their product verifies 
the constraint Eq.~(\ref{modulus}) until for large $x$ the approximation 
breaks down.


\end{document}